\documentclass[prb,twocolumn,floatfix,showpacs,preprintnumbers,superscriptaddress]{revtex4}
\usepackage{amsmath,amsfonts}
\usepackage{graphicx}
\usepackage{dcolumn}

\begin{document}

\title{Effects of interorbital hopping on orbital fluctuations
and metal-insulator transitions: Extended linearized dynamical
mean-field theory}
\author{Yun Song}
\affiliation{
    Department of Physics, Beijing Normal University, \\
    Beijing 100875, China.
}
\author{Liang-Jian Zou}
\affiliation{ Key Laboratory of Material Physics,
       Institute of Solid State Physics, Chinese Academy of Sciences, \\
       P. O. Box 1129, Hefei 230031, China
}

\date{\today}

\begin{abstract}

    We study the effects of interorbital hopping on orbital
fluctuations and Mott-Hubbard metal-insulator transition (MIT) in
the two-orbital Hubbard model within the extended linearized
dynamical mean-field theory. By mapping the model onto an effective
model with different bandwidths through the canonical
transformation, we find that at half-filling, the increases of the
interorbital Coulomb interaction $U^{\prime}$ and the Hund's
coupling $J$ drive the MIT, and the critical $J_{c}$ for MIT
increases with the lift of the inter-orbital hopping integral
$t_{ab}$. Meanwhile at quarter filling and in the strong correlation
regime, the system without $t_{ab}$ exhibits MIT with the decreasing
of $J$, and favors the orbital liquid ground state. However, the
system transits from metal to insulator with the increasing of
t$_{ab}$, accompanied with the rising of the orbital order
parameter. These results show the important role of the interorbital
hopping in the orbital fluctuation and orbital ordering.
\end{abstract}

\pacs{71.30.+h, 71.70.Ej, 71.27.+a, 71.20.-b}

\maketitle

\section{Introduction} 

In recent years, significant interest has been attracted to the
problem of orbital ordering and its effect on Mott-Hubbard
metal-insulator transition (MIT) of some transition-metal oxides
\cite{Imada}. As a result of the addition of orbital degree of
freedom, some strongly correlated 3d transition-metal oxides
exhibit very rich and complicated phase diagrams
\cite{Tokura,Carter,Maeno,Nakatsuji}. In these strongly correlated
systems with charge, spin and orbital degrees of freedom, various
orbital, magnetic and charge orders coexist or compete with each
other. Since many different quantum phases are almost degenerate,
a little change in orbital configuration usually leads to quantum
phase transition. This has been a hot topic in recent years since
the long-range orbital ordered ground state is found in colossal
magnetoresistive parent material $LaMnO_{3}$ \cite{Ishihara} and
the typical MIT system $V_{2}O_{3}$ \cite{Paolasini}. The quantum
fluctuations are important in these compounds, and it comes to be
the key issue to realize what role the orbital degree of freedom
plays in driving the quantum phase transitions. The conventional
self-consistent mean-field approach underestimates the effect of
the orbital quantum fluctuations, the dynamical
mean-field theory (DMFT) \cite{Georges} is an useful tool to
unveil the role of the quantum fluctuations on the Mott-Hubbard
MIT in these strongly correlated systems.

The DMFT, which is based on a mapping of lattice models onto
effective quantum impurity models subject to a self-consistency
condition, has been very successful in describing many aspects of
strongly correlated electron systems \cite{Georges}.
Many well-established numerical methods have been applied to solve the
effective quantum impurity model, such as the quantum Monte Carlo
\cite{Jarrell,Rozenberg,Georges2}, the exact diagonalization
\cite{Caffarel,Rozenberg2,Si}, and the renormalization-group
method \cite{BullaNRG1,BullaNRG2,BullaNRG3}. Although these
methods are successful in studying the MIT in single-band Hubbard
model, computational limits are met in applying the DMFT methods
to more complicated multi-band models.
Recently Bulla and Potthoff \cite{BullaLDMFT} proposed the
linearized DMFT (LDMFT) to solve the effective impurity Anderson
model with one impurity site and one bath site only, and this
ansatz shows minimum computational effort in dealing with the MIT
of Hubbard model. However, the LDMFT is bound to the critical
point in parameter space. Then Potthoff \cite{Potthoff} extended
this method (ELDMFT), which is able to
access the entire parameter space. The ELDMFT reduces to the LDMFT
at half-filling and at $U=U_{c}$.
The ELDMFT simplifies the procedure
of DMFT by linearizing the self-consistent equations in the low
and high energy regions, and can be easily applied to study some
more complicated models which are hard to be treated by the full
DMFT. The results obtained for the Hubbard model are in good
agreement with the other numerical techniques \cite{Potthoff}.

The minimal model to describe the Mott-Hubbard MIT including the
orbital degree of freedom is the two-orbital degenerate Hubbard
model. This model has been investigated with DMFT by several groups
\cite{Rozenberg3,Ono,Koga,Liebsch,Sato,Pruschke}: Rozenberg
\cite{Rozenberg3} found that the successive MIT occurs when the
electron occupation number varies from n=0 to 4; Koga $et$ $al.$
\cite{Koga} showed that two MITs may occur for the two subbands at
different Coulomb interaction; and Pruschke $et$ $al.$
\cite{Pruschke} recently demonstrated that the Hund's coupling plays
an important role on the MIT. Whereas most of these studies
neglected the charge transfer between the two different orbitals,
the inter-orbital charge fluctuation was prohibited and the effect
of orbital fluctuation was greatly underestimated. This paper is to
understand the orbital physics in multi-band Hubbard model at finite
temperatures, especially for the influences of the interorbital
hopping and Coulomb interaction on the orbital fluctuation, orbital
ordering and the MIT in the systems with various electron fillings.

In what follows, we study the two-orbital Hubbard model to explore
the role of orbital degrees of freedom and the interorbital
hopping, Hund's coupling and the interorbital Coulomb correlation
on MIT both in the half and quarter filling cases, respectively.
We find the Hund's coupling plays different roles in the half- and
quarter-filling systems; the interorbital Coulomb interaction
further splits the two Hubbard bands and drives the system from
metal to insulator at half filling; the orbital fluctuation in the
presence of interorbital (off-diagonal) orbital hopping is so
strong that the metallic phase is stable at quarter filling; also
the interorbital hopping may establish the orbital order and
drives the system to the insulating phase. The paper is organized
as follows: in Sec.II, we transform the two-orbital Hubbard model
with interorbital hopping into an effective model with different
bandwidth by the canonical transformation, which is easily mapped
into an impurity model in the framework of the DMFT. Then we
describe the extended linearized DMFT approach in Sec.III. Our
numerical results for the systems with half and quarter filling at
finite temperatures are presented in Sec.IV. And Sec.V is devoted
for the discussion and summary.

\section{The Two-orbital Hubbard Model} 

The two-orbital Hubbard model consists of the kinetic energy and
the Coulomb potential parts,
\begin{eqnarray}
H=H_{t}+H_{U}
\end{eqnarray}
with
\begin{eqnarray}
H_{t}&=&-\sum_{\langle ij\rangle} \sum_{ll^{\prime}\sigma}
t_{ll^{\prime}} C^{+}_{il\sigma}C_{jl^{\prime}\sigma}
     -\mu\sum_{il\sigma}C^{+}_{il\sigma}C_{il\sigma}
\nonumber \\
 H_{U} &=& U\sum_{il}n_{il\uparrow}n_{il\downarrow}
         +\frac{J}{2}\sum_{i,l\neq l^{\prime},
           \sigma}C^{+}_{il\sigma}C^{+}_{il\bar{\sigma}}
          C_{il^{\prime}\bar{\sigma}}C_{il^{\prime}\sigma}
\nonumber\\
         &+&\sum_{i,l\neq l^{\prime},\sigma
            \sigma^{\prime}}(\frac{U^{\prime}}{2}
         n_{il\sigma}n_{il^{\prime}\sigma^{\prime}} +\frac{J}{2}
         C^{+}_{il\sigma} C^{+}_{il^{\prime}\sigma^{\prime}}
         C_{il\sigma^{\prime}}C_{il^{\prime}\sigma}).
\nonumber\\
\end{eqnarray}
Here $l$ ($l^{\prime}$) = $a$ or $b$, and $t_{aa}$ ($t_{bb}$) and
$t_{ab}$ denote the nearest-neighbor intraorbital and interorbital
hopping integrals, respectively; $U$ describes the on-site
intraorbital repulsion between electrons and $U^{\prime}$ the
on-site interorbital interaction; and $J$ denotes the Hund's
coupling. 
There are two terms with $J$ in Eq. (2). The first one corresponds
to the spin exchange of the two intraorbital electrons, and the
second one to the orbital exchange between two different orbital
electrons. Such a full Hamiltonian keeps the rotation invariance of
the Hamiltonian (1) in the spin and orbital spaces. Usually the spin
exchange parameter is slightly different from the orbital one, while
for clarity we assume the two parameters are the same. 
For an isolated ion we have the rotation invariant relation
$U-U^{\prime}=2J$; in
some crystals, however, the direct Coulomb interactions $U$ and
$U^{\prime}$ are modified by the screening effect, and the
parameter $U^{\prime}$ and $J$ should be considered as independent
parameters \cite{Khomskii}. Here we consider these two cases,
respectively; and for clarity, we concentrate on the orbital
fluctuation, and fix the spin configuration as the paramagnetic
phase,
that is, we look for dynamic mean-field solutions that do not
break the spin rotational symmetry in this paper.

   For such multi-orbital systems with off-diagonal orbital hopping,
it is hard to map the Hamiltonian (1) onto the single impurity
effective model. We introduce two fermions $\alpha^{\dag}$ and
$\beta^{\dag}$ via a canonical transformation, and the kinetic
terms become of orbital diagonal in this quasi-particle
representation.
The canonical transformation is expressed as:
\begin{eqnarray}
C_{ia\sigma}&=&u\alpha_{i\sigma}+v\beta_{i\sigma} \nonumber\\
C_{ib\sigma}&=&-v\alpha_{i\sigma}+u\beta_{i\sigma},
\end{eqnarray}
with
\begin{eqnarray}
  u^{2} = \frac{1}{2} [1+\sqrt{1-
          \frac{t_{ab}^{2}}{t_{ab}^{2}+(t_{aa}-t_{bb})^2}}],
\nonumber\\
  v^{2} = \frac{1}{2}[1-\sqrt{1-
          \frac{t_{ab}^{2}}{t_{ab}^{2}+(t_{aa}-t_{bb})^2}}].
\end{eqnarray}
Here $\alpha^{\dag}$ and $\beta^{\dag}$ are the new fermion
operators, respectively. In this situation the twofold degenerate
orbital degree of freedom is converted into two different kinds of
fermions, and the orbital correlation is transformed into the
particle-particle correlation.
By Eq.(3), the kinetic energy $H_{t}$ in Eq.(1) is expressed as,
\begin{eqnarray}
   H_{t}&=&-t_{\alpha}\sum_{<i,j>,\sigma} \alpha^{+}_{i\sigma}
           \alpha_{j\sigma} -t_{\beta}\sum_{<i,j>,\sigma}
           \beta^{+}_{i\sigma}\beta_{j\sigma}
\nonumber\\
    &&-\mu\sum_{i,\sigma}(\alpha^{+}_{i\sigma} \alpha_{i\sigma} +
      \beta^{+}_{i\sigma} \beta_{i\sigma}),
\end{eqnarray}
with $t_{\alpha}=t_{aa}u^{2}+t_{bb}v^{2}-t_{ab}uv$ and
$t_{\beta}=t_{aa}v^{2}+t_{bb}u^{2}+t_{ab}uv$. In addition, we
transform the potential energy $H_{U}$ via the canonical
transformation Eq.(3), and obtain an effective two-orbital
Hamiltonian with different bandwidths so long as t$_{ab}$ $\neq$ 0.
Through the canonical transformation the two-orbital fermions with
interorbital hopping are transferred into two new fermions without
interorbital hopping, which can be treated more easily in the DMFT
scheme as to the single Hubbard model. On the other hand, the
on-site Coulomb energy and the Hund's coupling in H$_{U}$ inevitably
become very complicated, which is not hard to treat in the DMFT
framework in the present fermion representation.

\section{The extended linearized DMFT} 

In the DMFT, the above effective lattice model can be mapped into
an impurity model with two orbitals,
\begin{eqnarray}
 H_{imp}&=&\sum_{l\gamma\sigma}\epsilon_{l\gamma\sigma}
   a^{+}_{l\gamma\sigma}a_{l\gamma\sigma} -\mu\sum_{\sigma}
   \{\alpha^{+}_{0\sigma} \alpha_{0\sigma}
  +\beta^{+}_{0\sigma}\beta_{0\sigma}\} \nonumber
\\
&+& \sum_{l,\sigma} [ V^{\alpha}_{l\sigma}
   (a^{+}_{la\sigma}\alpha_{0\sigma}+h.c)
 + V^{\beta}_{l\sigma} (a^{+}_{lb\sigma}\beta_{0\sigma}+h.c)]
\nonumber \\
&+& H_{U}^{0}(\alpha, \beta),
\end{eqnarray}
where $\alpha_{0\sigma}$ (or $\beta_{0\sigma}$) annihilates a
fermion with spin $\sigma$ at the impurity site $i=0$. The
impurity couples with the bath which is described by operators
($a^{+}_{l\gamma\sigma}, a_{l\gamma\sigma}$) with energy
$\epsilon_{l\gamma\sigma}$ via hybridization
$V^{\gamma}_{l\sigma}$ ($\gamma=\alpha$ or $\beta$). For this
model, the free ($U=U^{\prime}=J=0$) impurity Green's function
$G^{0}_{\sigma}(\gamma,\gamma^{\prime})=\langle\langle
\gamma_{0\sigma}; ~\gamma^{\prime +}_{0\sigma}\rangle\rangle$
($\gamma_{0\sigma}$ ($\gamma_{0\sigma}^{\prime}$) =
$\alpha_{0\sigma}$ or $\beta_{0\sigma}$) is a $2\times 2$ matrix,
and it can be obtained via,
\begin{eqnarray}
&&(G_{\sigma}^{0}(\omega))^{-1}=  ~~~~~~~~~~~~~~~~~~~~~~~~~~
 \nonumber\\
 & &  \left(
       \begin{array}{cc}
       \omega+\mu-\sum_{l\sigma}\frac{(V^{\alpha}_{l\sigma})^{2}}
             {\omega-\epsilon_{l\alpha\sigma}} & 0 \\
       0 &\omega +\mu-\sum_{l\sigma}\frac{(V^{\beta}_{l\sigma})^{2}}
             {\omega-\epsilon_{l\beta\sigma}}
       \end{array}
   \right)
\end{eqnarray}

The Green's function $G_{\sigma}(\omega)$ of the impurity model in
Eq.(6) is obtained by the exact diagonalization method
\cite{Georges}, and thus the self-energy matrix is directly
extracted by Dyson equation,
\begin{equation}
\Sigma_{\sigma}(\omega)=(G_{\sigma}^{0}(\omega))^{-1}-
(G_{\sigma}(\omega))^{-1}.
\end{equation}
Therefore, the on-site Green's function reads,
\begin{equation}
G_{\sigma}(\omega)=\int^{\infty}_{-\infty}d\epsilon
\frac{\rho_{0}(\epsilon)}{\omega+\mu-\epsilon-
\Sigma_{\sigma}(\omega)}.
\end{equation}
Here we choose the semicircle density of states (DOS)
$\rho_{0}(\epsilon)_{\alpha,\beta}=\sqrt{4t^{2}_{\alpha, \beta}-
\epsilon^{2}}/(2\pi t^{2}_{\alpha,\beta})$ with respect to the two
branch fermions. The self-consistent equations (8) and (9) are
satisfied for proper values of the parameters
$\epsilon_{l\gamma\sigma}$ and $V^{\gamma}_{l\sigma}$ of the
impurity model by iterative calculation.

The exact diagonalization method has been combined with DMFT, in
which finite orbitals are chosen to describe the
bath\cite{Georges}. This method is exactly only as the number of
the bath orbitals $N_{b}$=$\infty$. However, in the study of
single band Hubbard model, it has been found that when $N_{b} >$
5, the corresponding results are good enough. While for more
complicated models, such as the multi-orbital degenerate models or
large cluster "impurity" model, the corresponding orbital
parameter space is much larger than that of the single-band
Hubbard model, and the self-consistent iteration will consume huge
computation time. In order to obtain precise results within the
ability of implementations of DMFT, the ELDMFT method is
constructed by considering an effective impurity model with one
impurity site and one bath site only\cite{Potthoff}.
This method simplifies the procedure of DMFT by linearizing the
self-consistent equations in the low and high energy regions, and
two self-consistent conditions are introduced to fix the two bath
parameters $\epsilon$ and $V$. For example in the single-band
Hubbard model, Potthoff compared the high-frequency expansion of
the on-site lattice Green's function with the result of the exact
high-frequency expansion, and found that the electron fillings
given by the Green's functions and by the impurity model is
equal\cite{Potthoff},
\begin{eqnarray}
n_{imp}=n.
\end{eqnarray}
Eq. (10) is the first self-consistent equation for the ELDMFT. To
satisfy the DMFT self-consistent equation (Eq. (9)) in the low
frequencies and match the features of the central quasiparticle
peak, Potthoff introduced the second self-consistent
equation\cite{Potthoff},
\begin{eqnarray}
V^{2}=ZM_{2}^{(0)}.
\end{eqnarray}
with $M_{2}^0=\int d\epsilon \epsilon^{2} \rho_{0}(\epsilon)$. For
a metal, $Z$ denotes the quasi-particle weight,
which represents the single particle excitation near the Fermi surface
of the metal. As $Z$ approaches zero, the single particle excitation
vanishes, indicating the occurrence of MIT.
The results for the single-band Hubbard model are in good agreement
with other numerical techniques \cite{Potthoff}.

To study the two-orbital Hubbard model, we introduce four bath
parameters ($\epsilon_{\alpha}$, $\epsilon_{\beta}$, $V^{\alpha}$,
and $V^{\beta}$) to describe the paramagnetic phase of the bath,
and thus four self-consistent conditions are introduced to fix
them accordingly,
\begin{eqnarray}
&&n^{\gamma}_{imp}=n_{\gamma} \nonumber\\
&&V_{\gamma}^{2}=Z_{\gamma}M^{(0)}_{2\gamma}
\end{eqnarray}
with $M_{2\gamma}^0=\int d\epsilon \epsilon^{2}
\rho_{0}(\epsilon)_{\gamma}$, and $\gamma=\alpha$ or $\beta$.
These equations renormalize the high- and low-frequency regimes of
the original self-consistency conditions in an integral,
qualitative form, and are thus well motivated.

\section{Theoretical Results} 

In studying the orbital fluctuations and MIT of the two-orbital
Hubbard model at finite temperature by the ELDMFT method, we
choose the original diagonal hopping as the unit,
$t_{aa}=t_{bb}$=1, and mainly consider two different conditions:
one is the spin symmetric case with $2J=U-U^{\prime}$, and the
other one is the case with the parameters $J$ and $U^{\prime}$
being independent. Without special note being made, the
temperature in this paper is $k_{B}T=1/\beta=1/16$. We restrict
the present study to the paramagnetic phase, the other possible
magnetic ordered phases shall be encountered latter on. In the
following studies, we focus on the systems with half filling
($n=2$) and quarter filling ($n=1$), respectively.

\subsection{Half filling cases}

The two-orbital Hubbard model reduces to two independent single-band
Hubbard models at $t_{ab}=U^{\prime}=J=0$. For the symmetric case
with half filling $n=2$, the MIT is observed at the critical Coulomb
interaction $U_{c}\approx 6.0$, which is consistent with the
Potthoff's result \cite{Potthoff}.  We now consider the systems with
$t_{ab}=J=0$ and $U=4$, and study the effect of the interorbital
Coulomb interaction on the MIT and orbital fluctuations.

\begin{figure}
\begin{center}
\includegraphics[width=7cm]{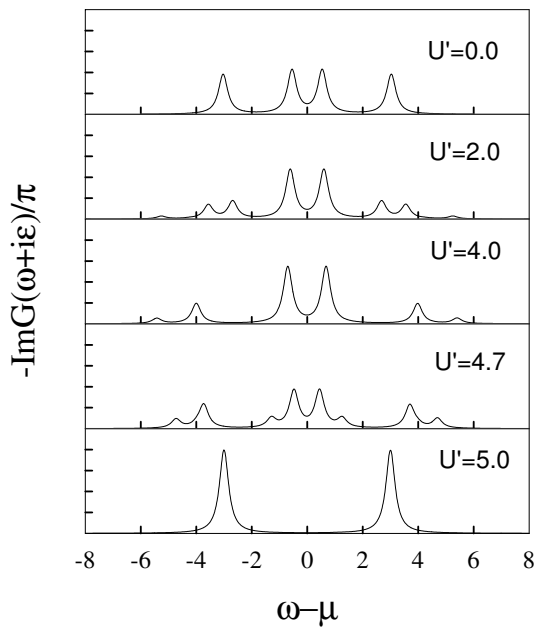}
\caption{Evolution of the DOS with the interorbital Coulomb
interaction in the half-filling systems. Theoretical parameters are
$U$=4, $J=0$ and $t_{ab}=0$.} \label{fig:denUpi}
\end{center}
\end{figure}

The DOS for the systems with $U^{\prime}$=0.0, 2.0, 4.0, 4.7 and 5.0
are shown in Fig.~\ref{fig:denUpi}.  At $U^{\prime}$=0 we find that
the quasiparticle spectra exhibit three distinct features: the
broad upper and lower Hubbard bands and the dominant quasiparticle
band near the Fermi energy, indicating that the system is a good
metal. This result resembles to that of the single-band Hubbard
model by the full DMFT \cite{Georges}. For the systems with finite
$U^{\prime}$, the spectral characters in the regime of
$U^{\prime}<U^{\prime}_{c}$ is quite different from that in
$U^{\prime}>U^{\prime}_{c}$, here $U^{\prime}_{c}$ denotes the MIT
critical point. In the regime of $U^{\prime}<U^{\prime}_{c}$, our
results clearly demonstrate that the interorbital correlations
remove the orbital degeneracy of the Hubbard bands and further split
the upper and lower Hubbard bands into four Hubbard subbands.
Approaching the critical point $U^{\prime}_{c}$, the DOS near the
Fermi surface decreases very quickly with the increase of
$U^{\prime}$, as shown in Fig.~\ref{fig:denUpi}. Large $U^{\prime}$
suppresses the quasiparticle excitation near E$_{F}$, and as a
result it leads to the MIT at the critical point of
$U^{\prime}_{c}=4.75$. What we obtained is in agreement with the
previous results by Rozenberg \cite{Rozenberg3} and Koga $et$
$al.$\cite{Koga}, confirming the validity and the reliability of the
present ELDMFT method.

   To learn the effects of the orbital fluctuations more better, we
also study the influence of thermal fluctuation on MIT at low
temperature $\beta=32$ and high temperature $\beta=$ 8 by keeping
the other parameters fixed. Considerable influence of the thermal
fluctuations are found, and the corresponding upper critical points
of MIT are $U^{\prime}_{c}$=5.0 for $\beta=32$ and
$U^{\prime}_{c}$=4.45 for $\beta=8$, respectively. We do not
consider a lower critical point of MIT with $U^{\prime}>U$,
presumptively unphysical. Obviously, the interorbital interaction
drives the MIT at high temperature more easier than at low
temperature, since the combination of the quantum and the thermal
fluctuations drives the paramagnetic metal transition to the
paramagnetic insulator more easily.

\begin{figure}
\begin{center}
\includegraphics[width=6.5cm]{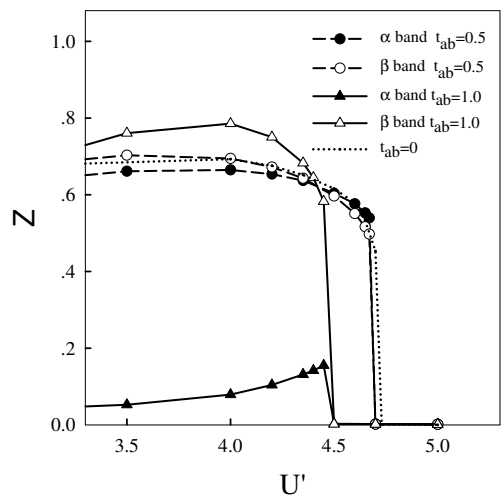}
\caption{Quasiparticle weights $Z$ (dotted line), $Z_{\alpha}$
(filled symbols) and $Z_{\beta}$ (open symbols) as the function of
$U^{\prime}$ at half filling with different interorbital hoping
$t_{ab}$.  The other parameters are: U=4 and J=0. }
\label{fig:ztpi}
\end{center}
\end{figure}

   Next we study the influence of interorbital hopping on the orbital
fluctuations and the MIT in the system with half filling. The
dependences of the quasiparticle weights on the interorbital
interaction in the cases with $t_{ab}=0.5$ and $t_{ab}=1.0$ are
shown in Fig.~\ref{fig:ztpi}. In Fig.~\ref{fig:ztpi}, the two dashed
lines with open and filled circles represent the quasiparticle
weight $Z$ of the effective orbitals $\alpha$ and $\beta$ as
$t_{ab}=0.5$ respectively, while the two solid lines with open and
filled triangles are the corresponding results for $t_{ab}=1.0$. For
the convenience of comparison, the result of the degenerate bands as
$t_{ab}=0$ (dotted line) is also shown. Firstly, we find that the
difference between the two effective bands $\alpha$ and $\beta$ is
significant as the interorbital hopping approaches the intraorbital
hopping. In comparison, the quasiparticle weights of the $\alpha$-
and $\beta$-orbitals are identical for the system without
off-diagonal hopping. In the presence of the interorbital hopping,
the quasiparticle weight of the $\beta$-band is larger than that of
the system with $t_{ab}=0$ in the metallic phase, while the
$\alpha$-band is smaller. As shown in Fig.~\ref{fig:ztpi}, the
contribution of the $\alpha$-band to the metallic properties at
$t_{ab}=1.0$ is small, the properties of the system is mainly
controlled by the $\beta$-band, implying that the orbital excitation
and orbital fluctuation are large and dominant in the metallic
phase. Secondly, the MIT critical point of the system is
$U^{\prime}_{c}=4.5$ for $t_{ab}=1.0$, obviously smaller than
$U^{\prime}_{c}=4.7$ for $t_{ab}=0.5$. The interorbital hopping
broadens the bandwidth of $\beta$-band and narrows that of
$\alpha$-band, and more electrons transfer to the $\beta$-band with
the increase of $t_{ab}$. As we have obtained above, the
interorbital interaction suppresses the quasiparticle excitation
near Fermi surface in the cases without off-diagonal hopping.
Similarly, the interorbital hopping also depresses the quasiparticle
weight of the $\alpha$-band by narrowing its bandwidth.  As a
result, the MIT is more easier to happen when $t_{ab}$ becomes large
in the paramagnetic half-filling system.

     Apart from the effects of interorbital hopping and Coulomb
correlation on MIT, our calculations also show that the Hund's
coupling $J$ plays a very important role in MIT. We find that for
$t_{ab}$=0, the behaviors of $Z$ for the $J\neq$0 case are quilt
different from the $J=0$ case. There are two MIT critical points
with $J_{c}=0.55$ and $J^{\prime}_{c}=-0.25$ (unphysical) for the
spin rotation symmetry case $2J=U-U^{\prime}$, and the metallic
phase can exist only in a narrow region $J^{\prime}_{c}<J<J_{c}$,
which is in agreement with Pruschke and Bulla's result
\cite{Pruschke}. The large Hund's coupling $J$ suppresses spin flip
excitation, and as $J$ increases the exchange splitting pushes the
$\alpha$-bands to higher with respect to the $\beta$-bands,
resulting in the system to transition to the insulating phase.
Therefore the role of the Hund's coupling is similar to the on-site
Coulomb Interaction $U$, driving the MIT by preventing the electrons
from double occupation. Meanwhile, instead of excluding the double
occupation, the interorbital Coulomb correlation drives the MIT by
removing the orbital degeneracy of the two Hubbard bands.

\begin{figure}
\begin{center}
\includegraphics[width=8.5cm]{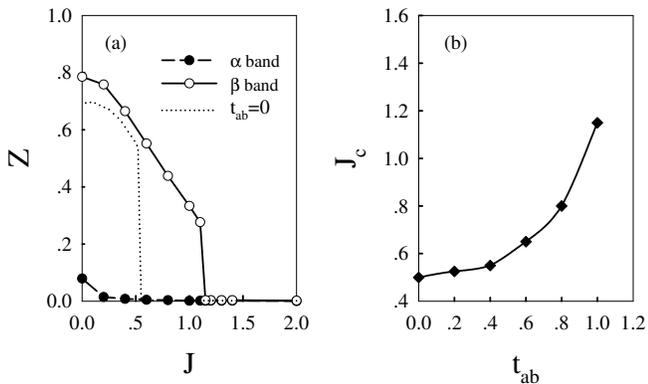}
\caption{(a) Dependence of quasiparticle weights $Z$ (dotted
line), $Z_{\alpha}$ (filled symbols) and $Z_{\beta}$ (open
symbols) as the function of J at $t_{ab}=1.0$, and (b) the
critical value $J_{c}$ as a function of interorbital hopping
integral $t_{ab}$ in the systems with $2J=U-U^{\prime}$.}
\label{fig:ZJcT}
\end{center}
\end{figure}

Further more we study the interplay between the interorbital hopping
and the Hund's coupling in driving the MIT. We show the dependence
of the quasiparticle weight on the Hund's coupling in the system
with $2J=U-U^{\prime}$ and $t_{ab}=1.0$ in Fig.~\ref{fig:ZJcT}(a).
For the convenience of comparison, the result of the degenerate
bands as $t_{ab}=0$ (dotted line) is also shown. When $J$ is small,
the system is metallic, and the quasiparticle weight of the
effective $\beta$-band is considerable larger than that of
$t_{ab}=0$. Since the hopping channels of the former is more than
the latter, the charge transfer between two different orbitals leads
to more considerable orbital quantum fluctuation than that in
$t_{ab}=0$. That means the presence of $t_{ab}$ broads the
conduction bandwidth and increases the DOS of the $\beta$-band, and
thus enhances the quasiparticle weight $Z_{\beta}$, as shown in
Fig.2. When $J$ becomes large, a lot of spin flip excitation is
suppressed, the system undergoes a MIT, correspondingly defines a
critical value J$_{c}$. The MIT critical point $J_{c}$ as a function
of the interorbital hopping is shown in Fig.~\ref{fig:ZJcT}(b).
Obviously the increase of the interorbital hopping enhance the
orbital fluctuations, and contributes more excitations near the
Fermi surface. As a result, more large J$_{c}$ are needed to drive
the MIT, and thus the MIT critical value $J_c$ increases with the
lift of the interorbital hopping.

\subsection{Quarter filling cases}

   The two-orbital Hubbard model with quarter filling
is appropriate to describe some important strongly correlated
materials with twofold degenerate orbital degree of freedom, such as
the hole-type compound $KCuF_{3}$ and electron-type compound
$LaNiO_{3}$. To our knowledge, the strongly correlated systems with
quarter filling have not been given enough attention by the
conventional DMFT yet. In this subsection we study the two-orbital
Hubbard model with quarter filling, we take large intraorbital
interactions ($U=8$ and 12), and assume the relationship
$U=U^{\prime}+2J$.

\begin{figure}
\begin{center}
\includegraphics[width=6.5cm]{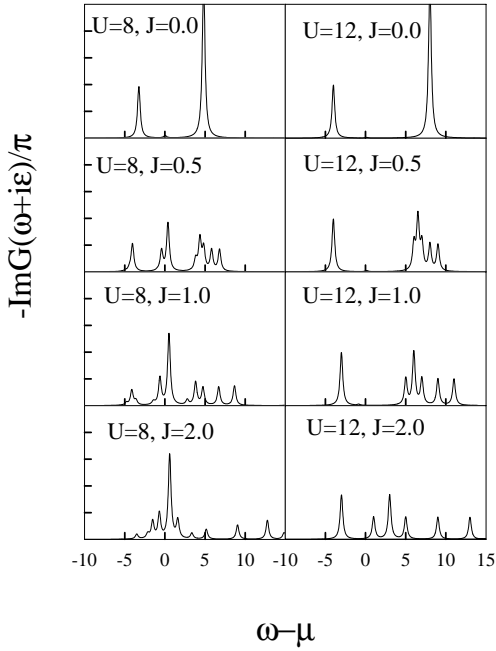}
\caption{Evolution of the DOS with the Hund's coupling J in the
quarter-filling systems with $U=8$ and $U=12$ at $t_{ab}$=0. }
\label{fig:D8&12}
\end{center}
\end{figure}

  First we consider the case with $t_{ab}=0$. The evolution of the DOS
with the Hund's coupling $J$ ($J=0.0$, 0.5, 1.0 and 2.0) for $U$=8
and 12 are shown in Fig.~\ref{fig:D8&12}. The Hund's coupling
strongly modifies the orbital subbands. At $J$=0, the presence of
the interorbital and intraorbital Coulomb interactions $U^{\prime}$
and $U$ removes nearly a half of the quasi-particle weight in the
lower Hubbard band to the upper Hubbard band,
and the lowest subband is thus almost filled up and the residue
quasiparticle weight near the Fermi surface is very small.
Therefore the system with $J=0$ at quarter filling is a bad metal
or an insulator.
In the system with $U$=8, the DOS at the Fermi level increases
gradually as $J$ increases; as $J > 0.5$ the system transits from
bad metallic or insulating phase to metallic phase.
Meanwhile, the increasing of Hund's coupling can further split the
energy bands into many subbands as shown in Fig.~\ref{fig:D8&12}. It
can be seen clearly that in the quarter-filling case, the influence
of the Hund's coupling $J$ on MIT is different to the half-filling
case: in the former the increase of the Hund's coupling indicates
the reduction of the interorbital repulsion $U^{\prime}$ and the
enhancement of the ferromagnetic fluctuation, 
therefore the metallic phase is in favorable;
however in the latter, the increase of $J$ leads the system to transit
from metal to insulator since the spin flip excitation is suppressed
and the quasiparticle weight near Fermi surface greatly declines.
On the other hand, for larger intraorbital Coulomb interaction
$U=12$, the system is always an insulator with the increasing of $J$
up to 2. As $J$ approaches to 2, the edge of the upper Hubbard bands
is gradually shifted to the Fermi energy, indicating the MIT will
occur at $J_{c}\approx$ 2.

\begin{figure}
\begin{center}
\includegraphics[width=7.5cm]{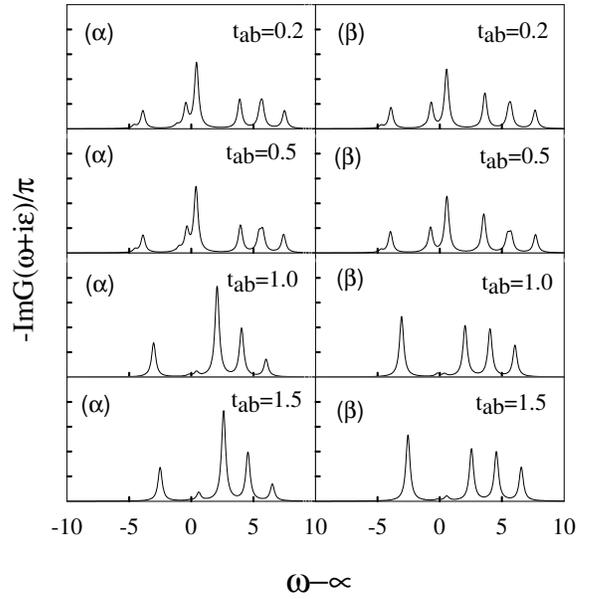}
\caption{Evolution of the DOS of $\alpha$- and $\beta$-bands with
the interorbital hopping $t_{ab}$ in the quarter filling systems
with $U=8$ and $J=1$. } \label{fig:GrTpi}
\end{center}
\end{figure}

   From Fig.4 one can see the striking difference of  the quasiparticle
weights between the systems with $U$ = 8 and with $U$=12. At $U=8$,
the quasiparticle weight $Z$ monotonously increases with the lift of
$J$, implying that the system is a good metal in the absence of the
interorbital hopping. However, for the system with $U=12$, the
quasiparticle weight $Z$ is very small, and the total weight is
0.031 as the Hund's coupling $J$ increases to 2. 
The double occupancy $D$, as one expects, decreases with the
increasing of the intraorbital repulsion. At $U=8$, the maximum
magnitude of the double occupancy is about $D_{m}=0.0012$. While at
$U=12$ the double occupancy $D_{m}$, about 0.0001, is negligible. 
Such strong single occupation and large Coulomb interaction in the
System with quarter filling suggests that the system with $U$=12 is
in the antiferromagnetic insulating phase, while the system with
$U$=8 is a paramagnetic or ferromagnetic metal. We will discuss
the cases with spin symmetry broken in a future paper.

  In the two-orbital Hubbard model at quarter filling, it is
very interesting how the interorbital hopping affects the MIT. Here
we mainly study the MIT in the systems with $U$=8 and $J$=1.0. The
evolution of the DOS of the $\alpha$- and $\beta$-bands with the
interorbital hopping is shown in Fig.~\ref{fig:GrTpi}. At small
$t_{ab}$ both $\alpha$- and $\beta$-bands cross the Fermi surface,
indicating that the system is metallic. Significantly different from
the half-filling cases, the further increase of the interorbital
hopping greatly reduces the quasiparticle states near the Fermi
surface for both the $\alpha$- and $\beta$-bands. One finds that in
the $\alpha$-band most states move to the high energy regime above
the Fermi level, while in the $\beta$-band, most states move to the
low energy regime below the Fermi level, thus the asymmetry between
the $\alpha$- and the $\beta$-band appears. And we find that as
$t_{ab}>1.0$, the DOS of the $\alpha$- and $\beta$-bands almost
vanish at the Fermi surface, implying an energy gap will open for
the two bands and MIT will occur. Since the electron kinetic energy
increases with t$_{ab}$, it is out of the expectation that the rise
of t$_{ab}$ leads to MIT. As we show in the following, this arises
from the formation of the orbital ordering in quarter-filling
systems with large interorbital hopping integrals.

\begin{figure}
\begin{center}
\includegraphics[width=6.5cm]{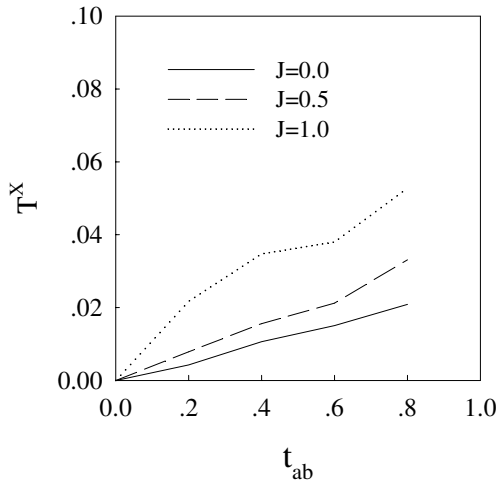}
\caption{Dependence of orbital order parameter $T^{x}$ on the
off-diagonal orbital hopping integral $t_{ab}$ in the
quarter-filling systems with $U=12$. \label{fig:Tx}}
\end{center}
\end{figure}

   In the presence of the strong Coulomb correlation, the system with
symmetric orbitals tends to break the orbital symmetry to lower
the ground state energy and form long-range orbital order. The
orbital order parameter, $<{\bf T}>$, is usually defined as the
average of the pseudospin operator  ${\bf T} = \sum_{ab}$
$C^{\dag}_{ia} {\bf \sigma}_{ab} C_{ib}$, with ${\bf \sigma}$
being the Pauli matrix, representing the orbital polarization of
the electrons occupation in the two orbitals. In the
quarter-filling system with $U$=12, we calculate the three
components of ${\bf T}$, $T^{x}$, $T^{y}$ and $T^{z}$, and the
results are shown in Fig.~\ref{fig:Tx}. From the preceding study,
we have found that in the absence of the interorbital hopping the
system with large interorbital interaction is in a para-orbital
phase \cite{Chen}. Switching on the off-diagonal or interorbital
hopping, we find that the symmetry of the orbital space is broken,
and the $x$-component of orbital order parameter $T^{x}$ is
finite, while the $z$-component $T^{z}$ is always zero. This
result is also confirmed by the self-consistent mean-field method
for the effective spin-orbital superexchange interaction in the
limit of large U \cite{Chen}. In addition, large interorbital
hopping results in strong orbital fluctuations, and large Hund's
coupling favors the ferro-orbital occupation, the competition of
these factors favors the orbital polarization in the
$x$-direction, leads to the orbital order parameter ${\bf T}$
lying in the $x$-direction. and $T^x$ increases with the
increasing of the interorbital hopping integrals and the Hund's
coupling $J$, as seen in Fig.~\ref{fig:Tx}.

In the absence of the interorbital hopping $t_{ab}$, the
interorbital Coulomb interaction $U^{\prime}$ and the Hund's
coupling $J$ adjust the relative positions of the four subbands;
though the electrons occupying one orbital configuration is
favorable in energy, no electron can transfer between these
orbitals, and the orbital occupations of the electrons in the two
orbital are the same, giving rising to orbital disordered phase. In
the presence of $t_{ab}$, the electrons can transfer between
orbitals and occupy the most favorable orbital configuration in
energy, and therefore the orbital ordering can establish, as we find
in the present results. 
It is worthy of noting that we only consider a possible ferro-orbital
phase, while more complicated orderings are out of reach from the
present single-impurity DMFT study. Further study will be
addressed in the future paper.

\section{Conclusions}

   From the proceeding study it is clear that the interorbital hopping
plays an important role in the MIT of the two-orbital degenerate
Hubbard models. The presence of the interorbital hopping not only
enhances the orbital fluctuations and quasi-particle excitation near
Fermi surface, but also leads to some unusual results. At half
filling, since more and more electrons transfer from one orbital to
the another, the critical value of the interorbital Coulomb
interaction $U^{\prime}_{c}$ for MIT decreases with the increase of
the interorbital hopping; while the critical value of the Hund's
coupling $J_{c}$ for MIT lifts with the increase of t$_{ab}$. In the
systems with $2J=U-U^{\prime}$, metallic phase can exist in a very
narrow region with the variation of J, since both intra- and
interorbital Coulomb interactions, $U$ and $U^{\prime}$, split the
Hubbard subbands and drive the MIT. At quarter filling, the increase
of $J$ may lead to insulator-metal transition, and this role in MIT
is completely contrary to that in the half-filling case; and the
increase of the interorbital hopping also may drive the MIT due to
strong orbital fluctuation, and leads to weak ferro-orbital ordering
in the system with large on-site Coulomb interaction.

  Meanwhile, the interorbital Coulomb correlation also drives the MIT by
expanding the separation of two orbital Hubbard subbands both in half-
and quarter-filling systems. In the present paper we mainly
concentrate our study on the paramagnetic and para/ferro-orbital phase,
little attention is paid to complicated spin and orbital orderings.
At present, it is not possible to compare the theoretical results
with the experimental results in candidate compounds KCuF$_{3}$
and LaNiO$_{3}$, since more complicated spin and orbital orders
are involved in these two realistic compounds, thus more than two
impurities must be considered in the study, beyond the
single-impurity DMFT theory.

Recently it receives great interest whether the wide and the narrow
Hubbard subbands exhibit MIT separately, i. e. the {\it
orbital-selective Mott transition} (OSMT) \cite{Koga,Medici} in
two-orbital Hubbard model. Obviously our results do not show that
there exist such Mott transition in the presence of interorbital
hoping $t_{ab} \neq 0$. While in the absence of t$_{ab}$, it is not
difficult to expect the existence of OSMT since the charge transfer
between orbitals is forbidden, and the role of the $a$-band is to
exert an effective potential on the $b$-band and to change the
chemical potential of the subband, therefore the MIT of the $a$-band
and the $b$-band occur separately. On the contrary for $t_{ab} \neq
0$ there exists charge transfer between the two orbitals, and the
MIT will occur at the same critical points for both bands. 
While for the system involving two subbands of different bandwidths
and omitting the interorbital hopping, as we find in the new
quasi-particle representation in Fig.3a, the OSMT seems happen as
the Hund's coupling is large enough. Further studies are deserved to
investigate the OSMT in the whole parameters space to determine the
phase diagram of the two-band Hubbard model with nondegenerate
subbands.

  In summary, utilizing the extended linearized DMFT, we find in the
two-orbital Hubbard models with the interorbital hopping and half
filling, the increase of Hund's coupling drives the MIT; on the
other hand, the quarter-filling two-orbital systems remain metallic
due to large orbital fluctuations. The systems with only
intraorbital hopping favors metallic and orbital liquid phase; as a
contrast, there exists the long-range orbital ordering in the
quarter-filling two-orbital systems with interorbital hopping and
large U.

\acknowledgements This work was supported by the grants from Beijing
Normal University, the NSFC of China and the Chinese Academy of
Sciences (CAS).



\begin{thebibliography}{10}

\bibitem{Imada}
M. Imada, A. Fujimori and Y. Tokura, {\it Rev. Mod. Phys.},
{\bf 70}, 1039 (1998).

\bibitem{Tokura}
Y. Tokura, A. Urushibara, Y. Moritomo, T. Arima, A. Asamitsu, G. Kido,
and N. Furukawa, {\it J. Phys. Soc. Jpn.} {\bf 63}, 3931 (1994).

\bibitem{Carter}
S. A. Carter, T. F. Rosenbaum, P. Metcalf, J. M. Honig, and J.
Spalek, {\it Phys. Rev.} {\bf B 48}, R16841 (1993).

\bibitem{Maeno}
Y. Maeno, H. Hashimoto. K. Yoshida. S. Nishizaki. T. Fujita, J. G.
Bednorz, and F. Lichtenberg, {\it Nature} (London) {\bf 372}, 532 (1994)

\bibitem{Nakatsuji}
S. Nakatsuji and Y. Maeno, {\it Phys. Rev. Lett.} {\bf 84}, 2666 (2000).

\bibitem{Ishihara}
S. Ishihara and S. Maekawa, {\it Phys. Rev. Lett.} {\bf 80}, 3799 (1998).

\bibitem{Paolasini}
L. Paolasini, C. Vettier, F. de Bergevin, F. Yakhou, D. Mannix, A. Stunault,
W. Neubeck, M. Altarelli, M. Fabrizio, P. A. Metcalf, and J. M. Honig,
{\it Phys. Rev. Lett.} {\bf 82}, 4719 (1999)

\bibitem{Georges}
A. Georges, G. Kotliar, W. Krauth and M. J. Rozenberg, {\it Rev.
Mod. Phys.} {\bf 68}, 13 (1996).

\bibitem{Jarrell}
M. Jarrell, {\it Phys. Rev. Lett.} {\bf 69}, 168 (1992)

\bibitem{Rozenberg}
M. J. Rozenberg, X. Y. Zhang, and G. Kotliar, {\it Phys. Rev. Lett.} {\bf 69}, 1236 (1992).

\bibitem{Georges2}
A. Georges and W. Krauth, {\it Phys. Rev. Lett.} {\bf 69}, 1240 (1992).

\bibitem{Caffarel}
M. Caffarel and W. Krauth, {\it Phys. Rev. Lett.} {\it 72}, 1545 (1994).

\bibitem{Rozenberg2}
M. J. Rozenberg, G. Moeller, and G. Kotliar, {\it Mod. Phys. Lett.} {\bf B 8}, 535 (1994).

\bibitem{Si}
Q. Si, M. J. Rozenberg, G. Kotliar, and A. E. Ruckenstein, {\it
Phys. Rev. Lett.} {\bf 72}, 2761 (1994).

\bibitem{BullaNRG1}
R. Bulla, A. C. Hewson, and Th. Pruschke, {\it J. Phys:} Condens. Matter {\bf 10}, 8365 (1998).

\bibitem{BullaNRG2}
R. Bulla, {\it Phys. Rev. Lett.} {\bf 83}, 136 (1999).

\bibitem{BullaNRG3}
R. Bulla, T. A. Costi, and D. Vollhardt, {\it Phys. Rev.} {\bf B
64}, 045103 (2001).

\bibitem{BullaLDMFT}
R. Bulla and M. Potthoff, {\it Eur. Phys. J. } {\bf B 13}, 257
(2000).

\bibitem{Potthoff}
M. Potthoff, {\it Phys. Rev.} {\bf B 64}, 165114 (2001)

\bibitem{Rozenberg3}
M. J. Rozenberg, {\it Phys. Rev.} {\bf B 55}, R4855 (1997); {\it
cond-mat/9612089}.

\bibitem{Ono}
Y. Ono, M. Potthoff, and R. Bulla, {\it Phys. Rev.} {\bf B 67}, 035119 (2003).

\bibitem{Koga}
A. Koga, Y. Imai, and N. Kawakami, {\it Phys. Rev.} {\bf B 66},
165107 (2002); A. Koga, N. Kawakami, T. M. Rice, and M. Sigrist,
{\it Phys. Rev. Lett.}  {\bf 92}, 216402 (2004).

\bibitem{Liebsch}
A. Liebsch, {\it Phys. Rev. Lett.} {\bf 91}, 226401 (2004).

\bibitem{Sato}
R. Sato, T. Ohashi, A. Koga, and N. Kawakami, {\it J. Phys. Soc.
Jpn.} {\bf 73}, 1864 (2004).

\bibitem{Pruschke}
T. Pruschke and R. Bulla, {\it cond-mat/0411186}.

\bibitem{Khomskii}
D. I. Khomskii and M. V. Mostovoy, {\it J. Phys}. {\bf A 36}, 9197
(2003); M. V. Mostovoy and D. I. Khomskii, {\it Phys. Rev. Lett.}
{\bf 92}, 167201 (2004).

\bibitem{Chen}
D.-M. Chen and Liang-Jian Zou, {\it unpublished}.

\bibitem{Medici}
L. de' Medici, A. Georges and S. Biermann, {\it cond-mat/0503764};
R. Arita and K. Held, {\it cond-mat/0503764}; A. Liebsch, {\it
cond-mat/0505393}.

\end{thebibliography}
\end{document}